\documentclass[preprint,1p]{elsarticle}
\usepackage[english]{babel}
\usepackage[usenames,dvipsnames]{color}
\usepackage{amssymb}
\usepackage{amsmath}
\usepackage{enumerate}
\usepackage{graphicx}
\usepackage{bm}
\usepackage{hyperref}
\usepackage{makeidx}

\begin{document}

\title{Tuning friction with composite hierarchical surfaces}

\author[to]{Gianluca Costagliola}
\ead{gcostagl@unito.it}
\author[to]{Federico Bosia} 
\ead{fbosia@unito.it}
\author[tn,qm,asi]{Nicola M. Pugno\corref{cor1}}
\ead{nicola.pugno@unitn.it}

\address[to]{ Department of Physics and Nanostructured Interfaces and Surfaces Centre, 
University of Torino, Via Pietro Giuria 1, 10125, 
Torino, Italy.}
\address[tn]{Laboratory of Bio-Inspired \& Graphene Nanomechanics, Department of Civil, 
Environmental and Mechanical Engineering, University of Trento, Via Mesiano, 77, 38123 Trento, 
Italy}
\address[qm]{ School of Engineering and Materials Science, Queen Mary University of 
London, Mile End Road, London E1 4NS, UK}
\address[asi]{Ket Lab, Edoardo Amaldi Foundation, Italian Space Agency, Via del Politecnico snc, 
00133 Rome, Italy}

\cortext[cor1]{Corresponding author}

\begin{abstract}

Macroscopic friction coefficients observed in experiments are the result of various types of 
complex multiscale interactions between sliding surfaces. Therefore, there are several ways to 
modify them depending on the physical phenomena involved. Recently, it has been demonstrated that 
surface structure, e.g. artificial patterning, can be used to tune frictional properties. In this 
paper, we show how the global friction coefficients can also be manipulated using composite 
surfaces with varying roughness or stiffness values, i.e. by combining geometrical features with 
the modification of local friction coefficients or stiffnesses. We show that a remarkable reduction 
of static friction can be achieved by introducing hierarchical arrangements of varying local 
roughness values, or by introducing controlled material stiffness variations.

\end{abstract}

\maketitle


\section{Introduction}

The constitutive laws of friction appear to be very simple at the macroscopic scale, indeed they 
were already formulated by Leonardo da Vinci, and later introduced in the context of 
classical mechanics with the so called Amonton's-Coulomb (AC) law: the friction force is 
proportional to the applied normal load and is independent of the apparent contact surface and of 
the sliding velocity \cite{persson}. The proportionality constants are called friction 
coefficients, which are different in the static and the dynamic sliding phase. Although some 
violations have been observed \cite{exp1}, this is a good approximate description of the 
macroscopic frictional force between two solid sliding surfaces \cite{popov}.

However, the origin of this behaviour turns out to be much more complicated, since friction 
coefficients are effective values, enclosing all the interactions occurring from atomic 
length scales, involving ``dry'' or chemical adhesion forces, to macroscopic scales, involving 
forces due to solid deformation and surface roughness. Moreover, friction coefficients are not a 
specific feature of the specific material, rather they are the result of the complex interplay 
between the contact surfaces occurring at various length scales in that material and involving 
different basic physical mechanisms \cite{nos1}\cite{bhu}. Thus, in order to modify the macroscopic 
emergent behaviour, one can intervene on the single mechanisms involved. 
For example, it is possible to modify the interactions at the microscopic level by means of
lubrication between surfaces, so that solid-solid molecular forces are switched to 
liquid-solid interactions and friction is reduced. At the macroscopic level, friction can 
be reduced by means of smoothing or polishing procedures, in order to remove surface
asperities hindering relative motion. Thus, problems related to friction, which is a complex 
multiscale phenomenon, can be addressed with different methods, from a practical and a
theoretical point of view \cite{rev}.

Another way to modify frictional properties is to manufacture sliding surfaces with artificial 
patterning, from micrometric to millimetric scales, e.g. grooves and pawls perpendicular to the 
direction of motion. The effects of these structures have been studied both numerically 
\cite{capoz3} and experimentally \cite{baum}\cite{maegcopi}, and recently their hierarchical 
arrangement has also been investigated by means of numerical simulations \cite{our}: results show 
that by changing the architecture of the contact surface only, the global static friction 
coefficients can be tuned without changing the chemical or physical properties of the material. 
This is because by exploiting patterning it is possible to modify mesoscopic features, i.e. the 
effective contact area and the stress concentrations  occurring in the static phase, providing a 
way to modify macroscopic friction coefficients.

In this paper, we show that this approach can be combined with the local variation of
friction coefficients, corresponding to a local change of material properties or of local 
surface roughness, in order to reduce static friction. We consider only roughness modifications 
occurring at the mesoscopic scale, using a statistical description based on a 
one-dimensional version of the spring-block model \cite{braun}. This approach allows to address the 
problem of friction in composite materials, which are widely used in practical applications 
\cite{comp1}-\cite{comp5} but whose frictional behavior is still scarcely studied from a  
theoretical and numerical point of view. Moreover, we consider local hierarchical arrangements of 
surface properties on different characteristic length scales. This allows us to highlight the main 
mechanisms taking place in the presence of different length scales, which could be exploited 
to design artificial surfaces with specific tribologic properties. 

Finally, we also consider a composite material with varying elastic properties, i.e. in which the 
elastic modulus is characterized by a linear grading. This can be found for example in 
functionally-graded composite materials, i.e. inhomogeneous materials whose physical 
properties are designed to vary stepwise or continuously \cite{suresh}\cite{gcomp} to manipulate 
global properties such as elasticity, thermal conductivity, hardness etc. These types of composite 
materials are widely adopted in practical applications, so that it is useful to investigate their 
frictional properties. A linear grading of elastic properties can be also combined with a local 
change of surface roughness in order to exploit both effects.

\clearpage

\section{Spring-block model}\label{sec_mod}

\begin{figure}[ht!]
\begin{center}
\includegraphics[scale=0.3]{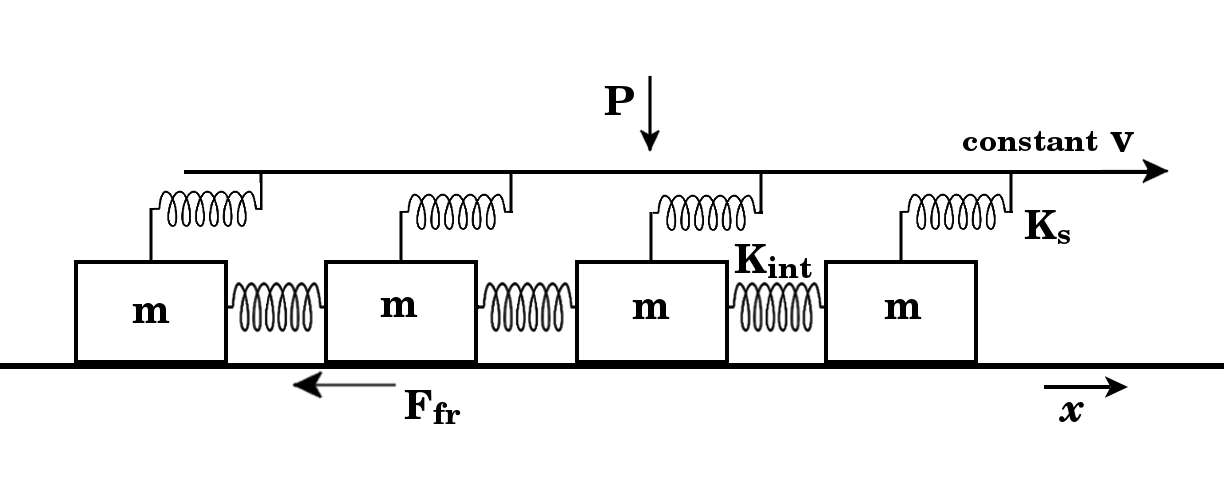}
\caption{\footnotesize Schematic of the spring-block model with the notation used in the text.}
\label{figmodel} 
\end{center}
\end{figure}

In order to study the effect of varying local friction coefficients on a surface, we 
adopt the one-dimensional spring-block model \cite{burr}\cite{equake1}, which is schematically 
represented in figure \ref{figmodel}: the material is discretized in $N$ blocks of mass $m$ along 
the direction of motion, connected by means of springs of stiffness $K_{int}$ and rest length 
$l_x$. Each block is also attached by means of shear springs of stiffness $K_s$ to a slider which 
is moving at constant velocity $v$. A normal pressure $P$ is applied uniformly to the surface, so 
that the same normal force is acting on all blocks. A viscous force with damping 
coefficient $\gamma$ in the underdamped regime is also added, in order to eliminate artificial 
block oscillations. Despite its simplicity, this model has already been used in many studies to 
investigate the frictional properties of elastic materials 
\cite{braun},\cite{maegawa}-\cite{capoz2}.

The blocks, representing a region of characteristic length $l_x$ on the surface of the 
material, are in contact with an infinitely rigid plane. Friction at the block scale is introduced 
through the classical AC friction force: each block is characterized by 
microscopic static and dynamic friction coefficients, respectively ${\mu_s}_i$, ${\mu_d}_i$, 
extracted from a Gaussian statistical distribution. In the following, we will drop the subscripts 
$s$ or $d$ of the friction coefficients every time the considerations apply to both the 
coefficients. 

This distribution does not necessarily represent the statistics of the contact points due to the 
surface roughness, rather it is a distribution of force thresholds for an elementary surface unit, 
used to provide an effective statistical description of the AC friction force at 
larger length scales than those relative to micro-scale phenomena. Though others can also be 
appropriate, the Gaussian distribution is a conventional choice that can be used to approximate any 
peaked distribution with parameters that are easily associated with the mean value and the 
standard deviation. The probability distribution is $p({\mu}_i) = (\sqrt{2\pi}\sigma)^{-1} 
\exp{[-({\mu}_i-(\mu)_m)^2/(2\sigma^2)]} $, where $(\mu)_m$ is the average 
microscopic coefficient and $\sigma$ is its standard deviation. This distribution is adopted for 
both the coefficients but with different parameters. 

The global friction coefficients, obtained from the sum of all the friction forces on the blocks, 
will be denoted with $M$, i.e. $(\mu)_M$. The global dynamic friction coefficient is calculated 
from the time average during the kinetic phase. The model does not include any roughness variation 
of value or other long term effects occurring after the onset of macroscopic sliding. Results 
regarding the dynamic friction are to be intended within the limits of this approximation. The 
global static friction coefficient is calculated from the maximum of the total friction force 
during the initial static phase, identified using the absolute maximum of the number of moving 
blocks, representing a macroscopic sliding event. In most cases, this coincides with the maximum of 
the total friction force over time.

In summary, the forces acting of each block are: the shear elastic force due to the slider 
uniform motion, $F_s = K_s (vt+l_i-x_i)$, where $x_i$ is the position of the block $i$ and $l_i$ is 
its starting rest position; the internal elastic restoring force between blocks $F_{int} = K_{int} 
(x_{i+1}+x_{i-1}-2x_i)$; the normal force $F_n=P \;l_x l_y$ and the viscous force $F_{d} = -m 
\gamma \dot{x_i}$; finally, the AC friction force $F_{fr}$: if the block $i$ is at rest, the 
friction force is equal and opposite to the resulting driving force, i.e. $F_{fr} = - (F_s + 
F_{int})$ up to the threshold $F_{fr} = {\mu_s}_i \; {F_n}$. When this limit is exceeded, a 
constant dynamic friction force opposes the motion, i.e. $F_{fr} = - {\mu_d}_i \; F_n$. Thus, the 
equation of the motion for the block $i$ along the sliding direction $x$ is obtained from 
Newton's law: $ m \ddot{x}_i = F_{int} + F_{s} -m \gamma \dot{x_i} + F_{fr}$.

The friction coefficients are fixed at the beginning of the simulation by extracting their 
values from the chosen distribution with a pseudo-random number generator. We have adopted a 
generator based on the Mersenne-Twister algorithm \cite{gent}. The overall system of 
ordinary differential equations can be solved numerically with a fourth-order Runge-Kutta 
algorithm with constant time step integration \cite{rk4}. Since the friction coefficients of the 
blocks are assigned after generating them with a pseudo-random number generator from the chosen 
distribution at each run, the final result of any observable consists on an average of various 
repetitions of the simulation. Usually, we assume an elementary integration time step $h=10^{-4}$ 
ms and we repeat the simulation about twenty times for statistical reliability.

The values of the parameters can be assigned by relating them to the macroscopic properties of the 
material, such as the Young's modulus $E$, the shear modulus $G$, the mass density $\rho$, the 
transversal dimensions $l_y$, $l_z$ and the total length $L_x=N l_x$. The mass is 
$m=\rho \; l_xl_yl_z$, the stiffnesses are $K_{int} = E\cdot(N-1) l_{y}l_{z}/ L_x $ and $K_s = 
G\cdot l_{y} l_{x}/l_{z}$. The stiffnesses are assumed constant for all the blocks, also in 
presence of different roughnesses, unless grading is explicitly introduced (see section 
\ref{sec_4}). This choice is made to reduce the number of free parameters of the model, but other 
formulations are equally valid (e.g. with constant friction coefficients and a statistical 
dispersion on the stiffnesses) and would not significantly affect the qualitative behaviour. We 
choose the global shear modulus as $G=5$ MPa, the Young's modulus $E=15$ MPa, the mass density 
$\rho=1.2$ g/cm$^3$, which are typical values for a rubber-like material with Poisson ratio 
$\nu=0.5$. 

The length $l_x$ is an arbitrary parameter representing the elementary discretization of the 
material and, consequently, the smallest surface feature that can be described in the model. We have 
fixed $l_x=0.05$ mm, corresponding to a size larger than micro-scale structures, like the surface 
roughness \cite{pers2}\cite{baum2} or microscopic patterns \cite{varenberg}-\cite{lee}. In any 
case, qualitative results are not affected by changing this parameter by an order of magnitude. The 
transversal lengths are fixed to $l_{z}=0.05$ mm, $l_y=1.0$ mm. The damping coefficient $\gamma$ is 
an arbitrary parameter which is tuned in the underdamped regime so that it is smaller than the 
characteristic frequencies of the system \cite{trom}. We fix $\gamma=100$ ms$^{-1}$, $N=480$, 
$v=0.05$ cm/s, $P = 0.1$ MPa. The microscopic friction coefficients are specified for each 
considered case.

\section{Friction on uniform surfaces}\label{sec_1}

The spring-block model described in section \ref{sec_mod} requires that microscopic friction 
coefficients be assigned to each block, extracting them from a statistical distribution, 
while the global friction coefficients are deduced by solving the equation of motion of 
the whole system. In this section, we investigate how the global friction coefficients are affected 
by these microscopic distributions of the friction coefficients, while other 
parameters are unchanged from section \ref{sec_mod}. 
This is useful to derive the behaviour of the model as a benchmark for the next sections, where more 
complex statistical distributions are introduced for the blocks. Here and in the next 
sections we will focus on the static friction coefficient, since the dynamic one has already been  
studied in \cite{our}.

\begin{figure}[ht!]
\begin{center}
\includegraphics[scale=0.45]{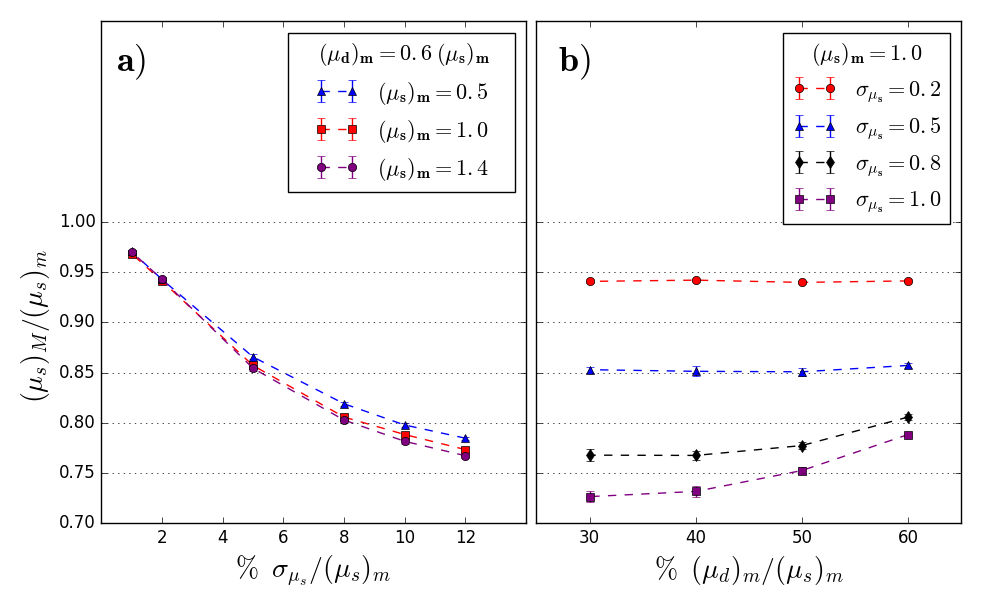}
\caption{\footnotesize a) Global static friction coefficients $(\mu_s)_M$ as a function of the 
statistical dispersion of the local ones $\sigma_{\mu_s}$ (with a fixed ratio between microscopic 
dynamic and static values $(\mu_d)_m$ and $(\mu_s)_m$, respectively). At first order the results 
do not depend on the values of the local static coefficients. 
b) Global static coefficients as a function the local dynamic coefficient (with a 
fixed local static one). For small variance values, the global static coefficient becomes 
insensitive to the local dynamic coefficient value. }
\label{fig_patt0} 
\end{center}
\end{figure}

In general, the numerical simulation of static friction is similar to that in a fracture mechanics 
problem \cite{fin}. The static friction coefficient distribution, corresponding to the
threshold forces for block motion, are analogous to the thresholds for breaking bonds in 
fiber bundle or lattice spring models \cite{small}\cite{lucas}. Hence, we expect the global 
friction coefficient to decrease with a wider static statistical distribution, since the presence of 
weaker elements can trigger avalanche ruptures leading to a macroscopic sliding event. This is 
confirmed by numerical simulations in figure \ref{fig_patt0}a:
for narrow statistical distributions, the relative reduction from $(\mu_s)_m$ to $(\mu_s)_M$ 
depends only on the ratio between microscopic static and dynamic coefficients. For a larger 
variance this is correct only as first approximation. This behaviour is taken as a reference for 
the cases considered in the following, when variations of the local coefficient distributions are 
introduced along then surface to model a spatially varying surface roughness.
Figure \ref{fig_patt0}b shows that the local dynamic coefficient influences the global static 
coefficient only for large variance values of the local static coefficients. This is because the 
macroscopic detachment phase (i.e. when some blocks are already in motion while others are still 
attached to the substrate) is longer with larger variances, so that the dynamic friction forces due 
to the moving blocks influence the total friction.

\clearpage

\section{Friction on variable-roughness patterned surfaces} \label{sec_2}

\begin{figure}[h!]
\begin{center}
\includegraphics[scale=0.45]{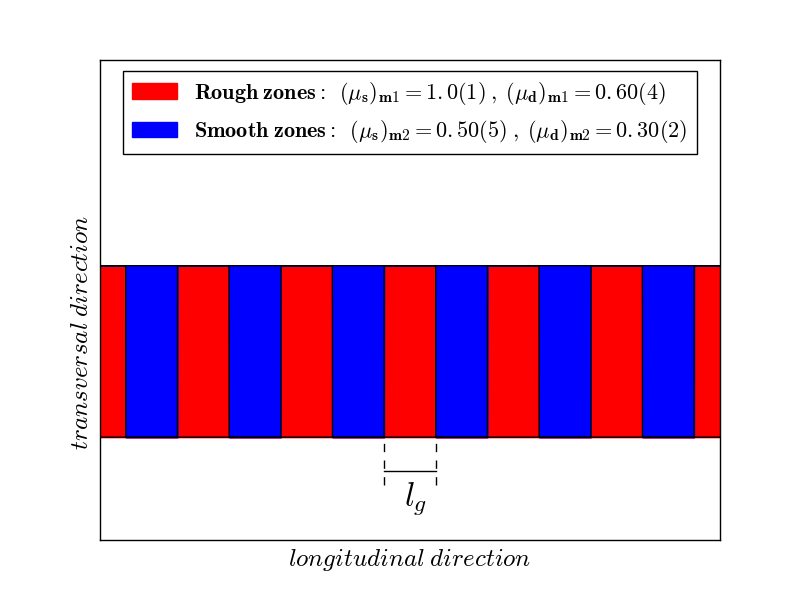}
\caption{\footnotesize Example of a pattern alternating rough and smooth zones of length $l_g$, 
with local friction coefficients respectively $(\mu)_{m1}$, $(\mu)_{m2}$. }
\label{fig1} 
\end{center}
\end{figure}

For the sake of simplicity, let us assume that our system can display two types of surface 
roughnesses (figure \ref{fig1}): for simplicity we will call them ``rough'' and ``smooth'' regions, 
although both have non negligible friction. In the rough regions of the surface the local 
friction coefficients are extracted from a probability distribution with $(\mu_s)_{m1} = 1.0(1)$ 
and $(\mu_d)_{m1} = 0.60(4)$, while in the smooth ones $(\mu_s)_{m2} = 0.50(5)$ and $(\mu_d)_{m2}  
= 0.30(2)$. 
The global friction coefficients for a uniform surface with these coefficients are $(\mu_s)_{M1} = 
0.788 (2)$, $(\mu_d)_{M1} = 0.616 (4)$ for rough regions, and $(\mu_s)_{M2} = 0.398 (2)$, 
$(\mu_d)_{M2} = 308(3)$ for smooth ones. As expected from section \ref{sec_1}, their ratio is 
about one half. 

Let us consider pattern of alternating rough and smooth regions with a characteristic length $l_g$, 
as depicted in figure \ref{fig1}. All other parameters of the system are fixed. In this 
configuration half of the surface is rough, and half is smooth. The number 
of blocks in a length $l_g$ is indicated as $n_g$, so that $l_g=n_g l_x$. Thus, the number of 
blocks in a rough zone $n_r$ or a smooth one $n_s$ are $n_r = n_s = N/(2 n_g)$.

The structure considered in figure \ref{fig1} is similar to a patterned surface with grooves, in 
which the friction coefficient is assumed to be zero. In this case, it is known 
both from numerical studies \cite{capoz3} and experimental results \cite{expat1}\cite{expat2} that 
static friction decreases with the width of the grooves. In our previous work \cite{our} we have 
also shown that this is due to the increase of the shear stress concentrations at the edge of the 
grooves. In the present situation, instead, the whole surface is in contact with the rigid 
substrate, but the mean value of the static friction coefficient varies periodically along the 
slider.

\begin{figure}[h!]
\begin{center}
\includegraphics[scale=0.5]{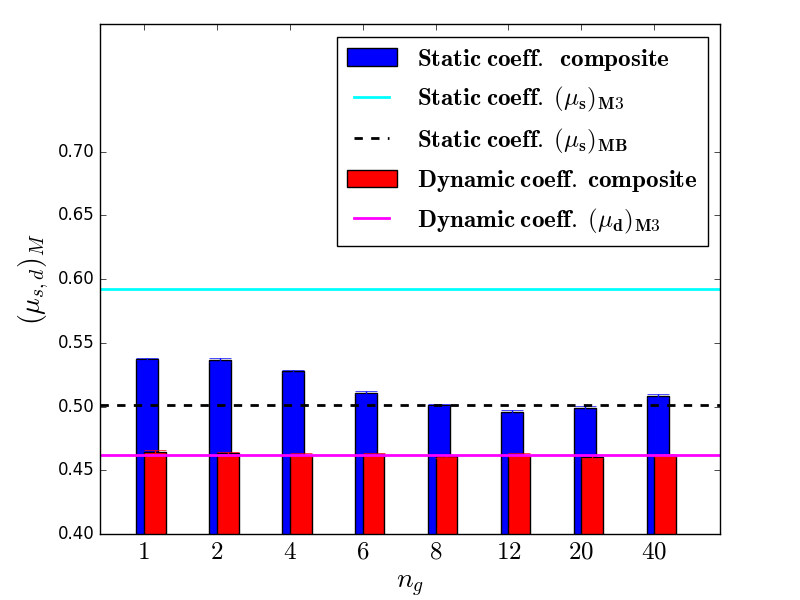}
\caption{\footnotesize Global friction coefficients as a function of the ratio between the widths 
of rough and smooth zones $n_g = l_g/l_x$. For comparison, we also show the global friction 
coefficients of a uniform surface, whose microscopic coefficients are extracted from a single 
Gaussian distribution with mean value corresponding to the arithmetic mean of the rough and smooth 
zones (the continuous lines, indicated with $(\mu)_{M3}$ in the text). The dotted line indicates the 
static coefficient $(\mu_s)_{MB}$ obtained with a bimodal Gaussian distribution.}
\label{fig_patt1} 
\end{center}
\end{figure}

In this case, the resulting global coefficient is expected to be included between the mean values 
of the two areas: $(\mu)_{M2} < {\mu}_M < (\mu)_{M1}$. In particular one could trivially think that 
the result for the global coefficient should be close to that obtained by setting as 
microscopic average $(\mu)_{m3}= ((\mu)_{m1}+(\mu)_{m2})/2 $, i.e. the arithmetic mean between the 
microscopic coefficients of the rough and smooth zones. Instead, as shown in figure 
\ref{fig_patt1}, this is true only for the dynamic coefficient, while the static one always  
displays a reduction  with respect to the average value. In particular, if we extract the 
microscopic coefficients of the blocks from a single Gaussian distribution corresponding to the 
arithmetic mean, $(\mu_s)_{m3} = 0.75(7)$ and $(\mu_d)_{m3} = 0.45(3)$, we obtain 
$(\mu_s)_{M3}=0.592(3)$ and $(\mu_d)_{M3}=0.462(1)$, while the coefficients obtained with a pattern 
of rough and smooth zones are always smaller. These results are consistent with those obtained 
with a multiscale version of the model, whose implementation is conceptually different, but applied 
to the same structure \cite{our2}.

This effect is due to the different variance between a single Gaussian distribution for all the 
surface and two separate Gaussian distributions: although the mean value is the same, the global 
static coefficient is reduced when a wider statistical dispersion is present, as shown in 
section \ref{sec_1}. If we extract all the local coefficients from a bimodal
Gaussian distribution (i.e. double peaked around $(\mu)_{m1}$ and $(\mu)_{m2}$), the result 
is $(\mu_s)_{MB} = 0.501 (2)$, i.e. twenty percent less than that of a single Gaussian. This 
configuration corresponds to a random arrangement of rough and smooth regions, as could be 
realized by a composite material, whose two component materials have different 
statistically distributed frictional properties. 

The resulting global friction properties are not dependent on statistical effects, since there is 
also an influence due to geometry: as shown in figure \ref{fig_patt1}, the size $l_g$ of the 
regions influences the global static coefficient similarly to what is observed with a patterning of 
grooves and pawls. The fundamental difference is that, in the present case, all blocks are always 
in contact, and the normal load is equally distributed along the whole surface, but a similar 
mechanism takes place: when the contact points of the surface in the smooth zones (typically with 
smaller threshold forces) begin to slide, they lead to an increase in the force exerted on the 
points still at rest in the rough zones, so that static friction is reduced with respect to 
$(\mu_s)_{M3}$ for any $l_g$. Moreover, the resulting global static coefficient can be either 
be tuned to be greater or smaller than $(\mu_s)_{MB}$, depending on the length $l_g$.

This example shows how the geometric organization of rough and smooth zones along the surface 
allows to modify static friction. In the following, we show that by combining this 
idea with a hierarchical structure, it is possible to obtain an even more consistent static 
friction reduction.

\section{Friction on surfaces with hierarchically patterned roughness} \label{sec_3}

\begin{figure}[h!]
\begin{center}
\includegraphics[scale=0.45]{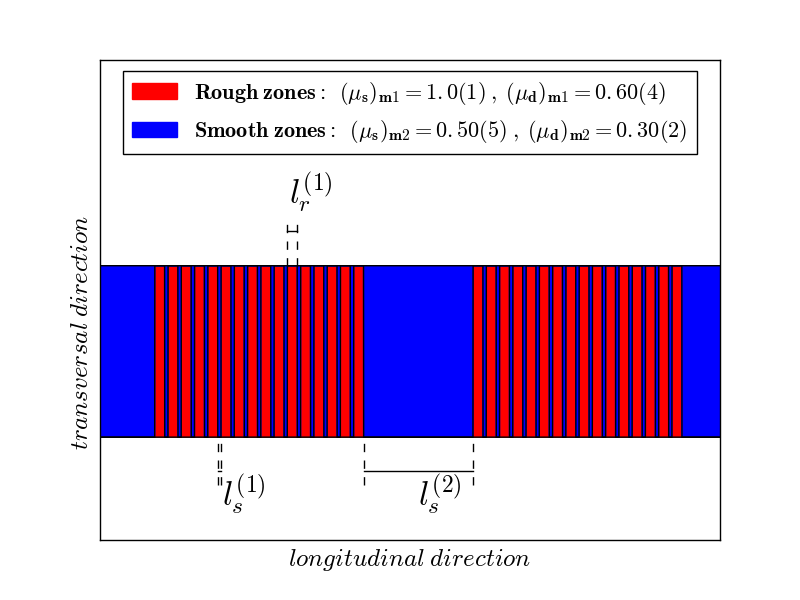}
\caption{\footnotesize Example of a surface with a hierarchical arrangement of smooth 
zones alternating with rough ones. The ratio between the sizes indicated in the 
figure are $l_s^{(2)}/l_s^{(1)} = 33$, $l_r^{(1)}/l_s^{(1)} = 3$, where the subscript $s$ denotes 
the smooth zones and $r$ the rough ones. The pattern is designed in such way that exactly half of 
the total surface is covered by rough zones and the other half by smooth zones.}
\label{fig2} 
\end{center}
\end{figure}

In this section, we investigate the effects on global friction coefficients induced by 
hierarchical organization of the regions with different roughnesses, as shown for example in 
figure \ref{fig2}. This configuration is hierarchical in the sense that there are two different 
length scales for the smooth regions (blue in the figure), that are included between the rough ones 
(red in the figure). To compare the results with those of section \ref{sec_2}, we choose 
a hierarchical pattern in which half of the overall surface area is smooth and the other is 
rough, so that the mean value of the distributions of the microscopic coefficients is 
still $(\mu)_{m3}$. 

\begin{figure}[h!]
\begin{center}
\includegraphics[scale=0.5]{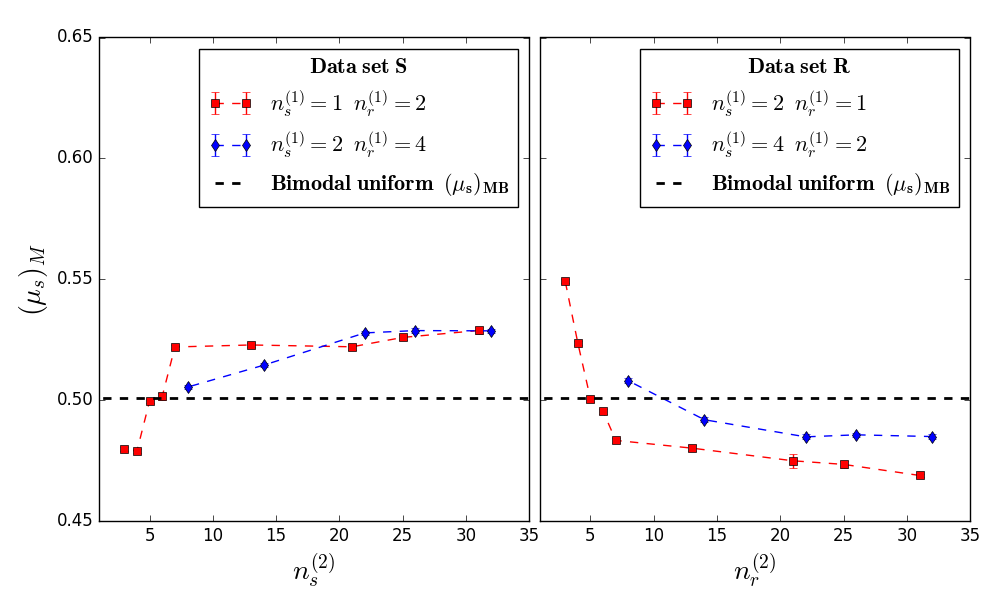}
\caption{\footnotesize Global static friction coefficients for hierarchical configurations such 
as the one shown in figure \ref{fig2}, as a function of the larger length scale, that is  
$n_s^{(2)}$ for data set S and  $n_r^{(2)}$ for data set R. The smaller length scales are reported 
in the legends. The dashed line indicates the case of a uniform surface with local friction 
coefficients $(\mu)_{MB}$ extracted from the bimodal Gaussian distribution.
}
\label{fig_patt2} 
\end{center}
\end{figure}

We identify such configurations by indicating the length of the smooth zones and the rough ones, 
respectively $l_s^{(i)}$ and $l_r^{(i)}$, ordered with the index $i$ increasing for the larger 
length scale. For example, the configuration shown in figure \ref{fig2} is characterized by the 
parameters $l_s^{(1)}$, $l_s^{(2)}$ and $l_r^{(1)}$, i.e. there are large smooth zones of size 
$l_s^{(2)}$, and then smaller rough and smooth zones of sizes $l_r^{(1)}$ and 
$l_r^{(1)}$, respectively. We can express these quantities using the adimensional ratios $n_s^{(i)} 
\equiv l_s^{(i)}/l_x$ and $n_r^{(i)} \equiv l_s^{(i)}/l_x$, representing the number of blocks for 
each region. Results for these configurations are shown in figure \ref{fig_patt2}.

A complementary configuration to that shown in figure \ref{fig2} can be obtained by exchanging the 
rough regions with the smooth ones, i.e. the subscript $s$ with $r$. In this case the 
statistics of the detachment thresholds for the configuration are exactly the same, but the 
geometry is different. In the case of a single level of patterning, as in section \ref{sec_2}, 
there is no effect by exchanging rough and smooth zones, while with a hierarchical arrangement the 
results are not symmetric and differ by up to ten percent. Since the statistics is the same, this 
effect is purely geometric. Thus, we have found a peculiar feature which can be obtained by means 
of hierarchical structures. We will call ``data set S'' that obtained with two length scales for 
the smooth regions (exactly as in figure \ref{fig2}) and ``data set R'' the complementary one, i.e. 
with two length scales for rough regions. 

\begin{figure}[ht]
\begin{center}
\includegraphics[scale=0.5]{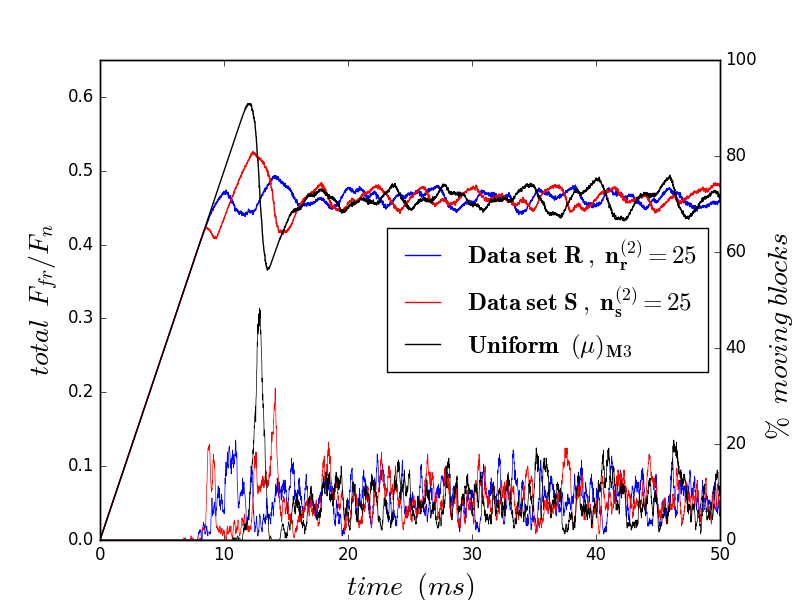}
\caption{\footnotesize Comparison of the total normalized friction force, as a function of 
time, between the data set S (case $n_s^{(1)}=2$, $n_r^{(1)}=1 $, $n_s^{(2)}=25 $ ), data set R 
(case $n_r^{(1)}=2$, $n_s^{(1)}=1 $, $n_r^{(2)}=25 $ ) and the uniform case with the same 
arithmetic mean of the local friction coefficient. The difference in the structure of the local 
friction coefficients of the illustrated cases causes a different qualitative transition between 
the static and dynamic sliding phase. On the right, axis the time evolution of the number of moving 
blocks is reported.
}
\label{fig_patt3} 
\end{center}
\end{figure}

As we can see in figure \ref{fig_patt2}, there are two different regimes leading to an increase or 
a decrease of the static friction with respect to the case of a uniform surface with local 
coefficients extracted from the bimodal Gaussian distribution: for data set S, static friction is 
greater for larger separations between length scales, i.e. for a large asymmetry between rough and 
smooth regions. Hence, configurations similar to data set R are preferable to reduce static 
friction and to ``flatten'' the transition between the static and kinetic phase. This is because of 
the interplay between microscopic degrees of freedom during the transition from static to kinetic 
friction, as shown in figure \ref{fig_patt3}, where the total friction force as a function of 
time is compared for both the data sets.

Thus, it is possible to tune the static friction by means of a hierarchical organization of zones 
with different roughnesses, and to obtain a friction coefficient close to the lower nominal limit 
$(\mu)_{M2}$, but with only half of the surface smoothed. From this we can conclude that to reduce 
the static friction of a material it is sufficient to smooth only part of the surface as long as it 
is in a ``smart'' way.

\clearpage

\section{Friction on surfaces with graded stiffness}\label{sec_4}

In this section, we investigate the modification of static friction due to the introduction 
of a linear grading of the elastic modulus, as occurs in a functionally graded composite material. 
We consider a linear increase (or decrease) of the elastic modulus along the longitudinal direction 
of the material, i.e. the sliding direction. In the spring-block model, this means that $K_{int}$ 
and $K_{s}$ depend on the block index $i$. In order to compare the results, the overall stiffness 
value $(K_{s})_{tot} \equiv \sum_i (K_s)_i$ is fixed, and similarly for $K_{int}$. Then, we 
introduce the relative maximum variation at the edges, namely $\Delta$, so that $\Delta=0.2$ means, 
for example, that for both the stiffnesses the maximum difference at the edge is twenty percent 
above/below their average. In symbols, $(K_s)_i = K_s (1+ \Delta (2 i/(N-1) -1) ) $ where $K_s$ is the 
value without grading, and the same holds for $K_{int}$.

\begin{figure}[h!]
\begin{center}
\includegraphics[scale=0.5]{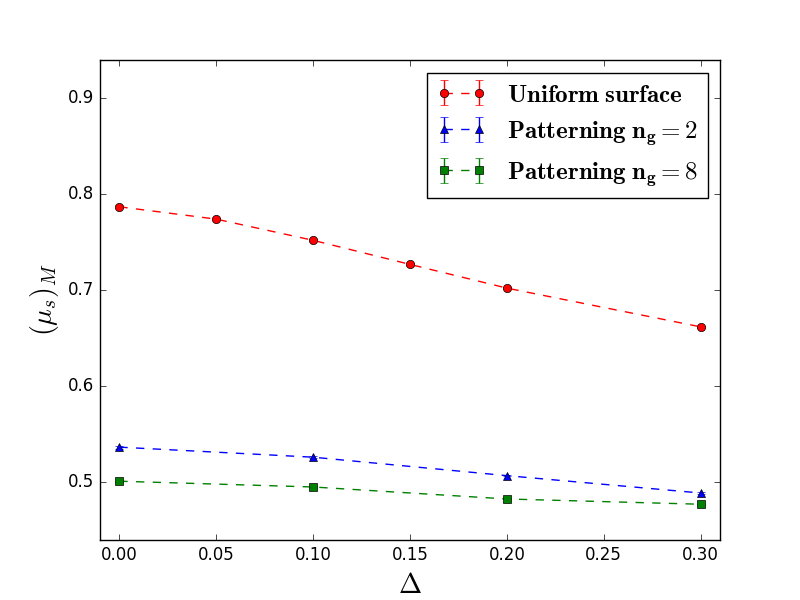}
\caption{\footnotesize Decrease of the macroscopic static friction coefficient as a function of the 
elastic modulus grading level for a uniform surface with microscopic 
coefficients $(\mu)_{m1}$ (red points), and for two cases of periodic patterning of the local 
roughness, as in section \ref{sec_2}, with $n_g=2$ (blue points) and $n_g=8$ (green points).
}
\label{fig_grad} 
\end{center}
\end{figure}

In the spring-block model, variations of $K_{int}$ turn out to be irrelevant, so that the 
effect can be studied by setting the grading only on the springs $K_s$. Results are shown in 
figure \ref{fig_grad}. In the presence of grading, the static friction coefficient is considerably 
reduced. The explanation for this is that  in this case the local rupture/sliding thresholds are 
exceeded sooner than in the case with no grading in the region where the stiffnesses are increased, 
so that an avalanche of ruptures is triggered in the neighbouring contact points, until the whole 
surface detaches. We observe no dependence on the orientation of the grading. Also the dynamic 
friction coefficient is left unchanged.

The exact amount of change of the static friction depends on the system parameters, but we may 
expect this effect on every configuration, because the grading always induces a stress distribution 
on the surface that favors avalanche phenomena. Indeed, a reduction of the 
static friction is also observed in the case of patterning of the local surface roughness. In these 
simulations, only a linear grading has been considered, but similar effects are expected with a 
generic functional shape. Thus, we have shown that a further reduction of the static 
friction can be obtained with a grading on the elastic properties of the material, i.e. in a 
composite material with a functionally-graded elastic modulus.

\clearpage

\section{Conclusions} 

In this paper, we have investigated by means of numerical simulations the variation of 
the friction coefficients of a material characterized by two distinct surface roughnesses and local 
friction coefficients, as found in composite materials, or in materials whose surfaces have 
different degrees of smoothing. For this purpose, we have adopted a one-dimensional version of the 
spring-block model, which is particularly appropriate for parametric studies on the frictional 
behaviour of a structured elastic material.

First, we have studied the effects due to statistical variations in surface roughness of a 
composite surface: the presence of a double-peaked distribution of the local friction 
coefficients implies that the variance is typically larger than for a single peak distribution, so 
that static friction is reduced. This effect also occurs without any surface patterning.

Secondly, we have evaluated the influence of the geometry on these composite systems. If the 
surface is divided into rough and smooth regions of the same size, the global static friction 
coefficient depends on their length scale, similarly to the case of a surface with a 
patterning of grooves and pawls. This effect is purely due to structure, since the statistics is 
the same, and the patterning can be used either to reduce the static friction or to increase it 
depending on the length scale of the different roughness zones. Thus, in order to considerably 
reduce static friction, it is sufficient to smooth only a part surface, as long as this is done in 
a ``smart'' way.

If instead we introduce different length scales for rough and smooth regions, i.e. we adopt a 
hierarchical organization of the zones with different roughnesses, we obtain opposite results 
depending on the ordering between rough and smooth zones. The surface is no longer symmetric under 
the exchange of the smooth and rough zones, and this has an influence on the global frictional 
behaviour. 
Thus, the geometric and multiscale arrangement is crucial to determine measurable variations 
of static friction, even when the statistical properties of the local coefficients are the same. 
This example suggests a possible mechanism for modifying the static friction properties of a 
surface by combining different mesoscopic roughness and geometric parameters. Additionally, it is 
possible to modify not only the numerical value of the static coefficient, but also the qualitative 
behaviour of the transition from static to dynamic friction.

Finally, we have considered a composite material with a graded elastic modulus by 
considering linearly varying stiffnesses in the spring-block model. Results show that this provides 
the possibility of a further reduction of the static friction, both in the case of a smooth surface 
and of patterning of the local roughness. 

All of these results can be relevant for a large number of applications where maximization 
or minimization of friction is crucial. One example could be the friction performance of vehicle 
tires that are typically produced in reinforced rubber composites with various levels of patterning 
or roughnesses. The large level of tunability of properties obtained exploiting composite material 
composition, stiffness, roughness and patterning provide an attractive way to reach desired properties, 
and the presented model a useful tool in the design of optimal solutions.

\clearpage 

\section*{Acknowledgments}
N.M.P. is supported by the European Research Council PoC 2015 ``Silkene'' 
No. 693670, by the European Commission H2020 under the Graphene Flagship 
Core 1 No. 696656 (WP14 ``Polymer Nanocomposites'') and FET Proactive 
``Neurofibres'' grant No. 732344. G.C. and F.B. are supported by H2020 FET 
Proactive ``Neurofibres'' grant No. 732344.

\end{document}